\documentclass[prd,amsmath,amssymb,floatfix,superscriptaddress,notitlepage,nofootinbib,preprintnumbers]{revtex4-1}
\usepackage{amsfonts,amssymb,amsmath,graphicx,color,bm,enumitem}
\definecolor{ultramarine}{rgb}{0.07, 0.04, 0.56}
\definecolor{cadmiumgreen}{rgb}{0.0, 0.42, 0.24}
\definecolor{indigo(dye)}{rgb}{0.0, 0.25, 0.42}
\usepackage[linktocpage=true]{hyperref}
\hypersetup{
colorlinks=true,
citecolor=ultramarine,
linkcolor=cadmiumgreen,
urlcolor=indigo(dye),
}

\newcommand{\f}[2]{\frac{#1}{#2}}  
\newcommand{\mk}[1]{\left( #1 \right)}  
\newcommand{\kk}[1]{\left[ #1 \right]}  
  
\newcommand{\be}{\begin{equation}}  
\newcommand{\ee}{\end{equation}}
\newcommand{\bem}{\begin{pmatrix}}
\newcommand{\eem}{\end{pmatrix}}
\newcommand{\Mpl}{M_{\rm Pl}}

\newcommand{\F}{\mathcal{F}}
\newcommand{\G}{\mathcal{G}}
\renewcommand{\H}{\mathcal{H}}

\renewcommand{\P}{\mathcal{P}}
\newcommand{\pa}{\partial}

\begin{document}

\preprint{YITP-18-104}

\title{
Stealth Schwarzschild solution in shift symmetry breaking theories
}

\author{Masato Minamitsuji}
\affiliation{Centro de Astrof\'{\i}sica e Gravita\c c\~ao  - CENTRA,
Departamento de F\'{\i}sica, Instituto Superior T\'ecnico - IST,
Universidade de Lisboa - UL, Av. Rovisco Pais 1, 1049-001 Lisboa, Portugal}

\author{Hayato Motohashi}
\affiliation{Center for Gravitational Physics, Yukawa Institute for Theoretical Physics, Kyoto University,\\ Kyoto 606-8502, Japan}

\begin{abstract}%%%%%%%%%%%%%%%%%%%%%%%%%%%%%%%%%%%%%%%%%
We find stealth Schwarzschild solutions with a nontrivial profile of the scalar field regular on the horizon in the Einstein gravity coupled to the scalar field with the k-essence and/or generalized cubic galileon terms, which is a subclass of the Horndeski theory breaking the shift symmetry, where the propagation speed of gravitational waves coincides with the speed of light.  After deriving sufficient conditions for the shift symmetry breaking theory to allow a general Ricci-flat metric solution with a nontrivial scalar field profile, we focus on the stealth Schwarzschild solution with the scalar field with or without time dependence. For the profile $\phi=\phi_0(r)$, we explicitly obtain two types of stealth Schwarzschild solutions, one of which is regular on the event horizon. The linear perturbation analysis clarifies that the kinetic term of the scalar mode identically vanishes, indicating that the scalar mode is strongly coupled. The absence of the kinetic term of the scalar mode in the quadratic action would inevitably arise for the stealth Schwarzschild solutions in the theory with a general scalar field profile depending only on the spatial coordinates. On the other hand, for the time-dependent scalar field profile, we clarify that there does not exist a stealth Schwarzschild solution in the shift symmetry breaking theories.
\end{abstract}
\keywords{Black hole; modified gravity}

\maketitle  

%%%%%%%%%%%%%%%%%%%%%%%%%%%%%%%%%%%%%%%%%  

\section{Introduction}
\label{sec1}

The recent data of gravitational waves (GWs) 
measured by the LIGO and Virgo Collaborations
from binary black hole (BH) mergers 
\cite{Abbott:2016blz,Abbott:2016nmj} 
and a binary neutron star merger
\cite{TheLIGOScientific:2017qsa}
with its optical counterparts \cite{GBM:2017lvd}
were highly consistent with the prediction of general relativity (GR).
With the latter data, 
the propagation speed of GWs traveling over cosmological distance
was shown to coincide with the speed of light down to the accuracy of order $10^{-15}$~\cite{Monitor:2017mdv}.
The future measurements of GWs with unprecedented accuracies
would be able to test modified gravity theories from different aspects.

Theories of modified gravity as an alternative to GR have attracted a lot of attention and  been extensively studied as a model to explain the late-time acceleration of the Universe~\cite{Clifton:2011jh,Berti:2015itd,Koyama:2015vza}. %,Heisenberg:2018vsk}.
The framework of scalar-tensor theories
which involve many representative modified gravity theories
has been extended 
to the Horndeski theory~\cite{Horndeski:1974wa,Nicolis:2008in,Deffayet:2009wt,Deffayet:2009mn,Deffayet:2011gz,Kobayashi:2011nu}
and even {\it beyond} it~\cite{Zumalacarregui:2013pma,Gleyzes:2014dya,Gleyzes:2014qga,Motohashi:2014opa,Langlois:2015cwa,Motohashi:2016ftl,Klein:2016aiq,BenAchour:2016fzp,Motohashi:2017eya,Motohashi:2018pxg}. 
%%%%%%%%%
The constraint on the propagation speed of GWs 
has ruled out some of these theories
as the origin of the late-time acceleration~\cite{Creminelli:2017sry,Sakstein:2017xjx,Ezquiaga:2017ekz,Baker:2017hug}
(See also Refs.~\cite{Lombriser:2015sxa,Lombriser:2016yzn}).
In the context of the Horndeski theory,
the theory which satisfies this bound 
is given by 
\begin{align}
\label{action}
 S=\int d^4x \sqrt{-g}
\left[
G_4 (\phi)
R 
+G_2 (\phi,X)
-G_3(\phi,X)\Box\phi
\right],
\end{align}
where
the indices $\mu,\nu,\cdots$ run the four-dimensional spacetime, 
$g_{\mu\nu}$ is the metric, 
$g={\rm det}(g_{\mu\nu})$,
$R$ is the scalar curvature associated with $g_{\mu\nu}$,
$\phi$ is the scalar field, 
$X:=-(1/2)g^{\mu\nu}\phi_\mu\phi_\nu$ is the canonical kinetic term of the scalar field,
$\phi_{\mu\nu\cdots\alpha}:= \nabla_\alpha\cdots \nabla_\nu\nabla_\mu \phi$
is the covariant derivative(s) of the scalar field
with respect to $g_{\mu\nu}$,
$G_4(\phi)$ is the function of $\phi$,
and 
$G_i (\phi, X)$ ($i=2,3$) are arbitrary functions of both $\phi$ and $X$.
%%%%%%

The models given by Eq.~\eqref{action}
also admit the propagation of the degrees of freedom of GWs,
i.e., the odd-parity mode and one of the even-parity modes,
with the speed of light
in the vicinity of static and spherically symmetric BHs~\cite{Kobayashi:2012kh,Kobayashi:2014wsa}.
In general, in the Horndeski theory,
the propagation speed of GWs would also be modified 
in the vicinity of localized gravitational sources
if the scalar field exists around them.
%%%%
Thus,
even if the scalar field is not the direct origin of the cosmic acceleration
of today,
the propagation speed of GWs may be modified
when they pass in the vicinity of them,
unless one considers the theory \eqref{action}.
Therefore,
the theory~\eqref{action} 
corresponds to the most conservative choice within the Horndeski theory
which sufficiently satisfies the current bound on the propagation speed of GWs,
assuming that the scalar field exists somewhere in the Universe.
These models will be the subject
for the future strong field tests on gravitation \cite{Berti:2015itd}.

In GR,
the Schwarzschild and Kerr BH solutions
which are solely determined 
by measuring the mass and angular momentum
\cite{Israel:1967wq,Carter:1971zc,Hawking:1971vc}
are known as the unique vacuum static and stationary solutions, respectively.
On the other hand, in general
scalar-tensor theories may possess BH solutions
different from the GR ones
\cite{Bocharova:1970skc,Bekenstein:1974sf,Kanti:1995vq,Radu:2005bp,Pani:2009wy,Anabalon:2009qt,Kleihaus:2011tg,Kolyvaris:2011fk,Anabalon:2012ta,Babichev:2013cya,Anabalon:2013oea,
Sotiriou:2013qea,Minamitsuji:2013ura,Herdeiro:2014goa,Ayzenberg:2014aka,Kobayashi:2014eva,Charmousis:2014zaa,Babichev:2016fbg,Babichev:2017guv,Erices:2017izj,Mukherjee:2017fqz}.
These theories
admit static or stationary BH solutions
different from the GR solutions
with nonconstant profiles of the scalar field \cite{Herdeiro:2015waa},
for instance,
in the Einstein-scalar-Gauss-Bonnet theories
\cite{Kanti:1995vq,Pani:2009wy,Kleihaus:2011tg,Sotiriou:2013qea,Ayzenberg:2014aka,Doneva:2017bvd,Silva:2017uqg,Antoniou:2017acq,Antoniou:2017hxj}
and in the Einstein-complex scalar theories \cite{Herdeiro:2014goa}.
However, it does not mean all the theories of modified gravity possess BH solutions different from GR.
There exist a particular class of theories whose equations of motion allow 
GR solutions with a constant profile of the scalar field~\cite{Motohashi:2018wdq}. 
Furthermore,
in some of these theories
the no-hair theorem was established;
i.e., they admit only the static and stationary BH solutions in GR
with a constant profile of the scalar field 
\cite{Chase,Bekenstein:1972ny,Hawking:1972qk,Bekenstein:1995un,Sotiriou:2011dz,Hui:2012qt,Graham:2014mda,Faraoni:2017ock}.

In the shift-symmetric Horndeski and beyond Horndeski theories,
the key assumptions that ensure the uniqueness of the 
GR BH solutions~\cite{Hui:2012qt,Babichev:2017guv} 
are
that
(i) the spacetime is static, spherically symmetric, and asymptotically flat,
(ii) the scalar field respects the symmetry of spacetime, i.e.,
solely the function of the radial coordinate for the case of the Schwarzschild solution,
(iii) the coupling functions and their derivatives are regular in the limit
of the vanishing canonical kinetic term,
and 
(iv) the canonical kinetic term dominates the other kinetic couplings
in the equations of motion at the spatial infinity.
The violation of (ii) by the the scalar field linearly depending on time yields
the so-called stealth Schwarzschild solution
with the nontrivial scalar field \cite{Babichev:2013cya}.
The violation of (iv)
by
the absence of the canonical kinetic term
gives the Kerr solution in the purely quartic Horndeski theory~\cite{Babichev:2017guv}.

On the other hand, 
the studies of 
the no-hair theorem and the stealth Schwarzschild solution in the shift-symmetry breaking 
Horndeski and beyond Horndeski theories
are still in the premature phase,
since once the shift symmetry is abandoned
all the coupling functions can be arbitrary functions 
of both the scalar field and the canonical kinetic term, 
which makes the analysis involved.
While general theories that allow GR solutions 
with a constant profile of the scalar field has been clarified~\cite{Motohashi:2018wdq},
the no-hair theorem has not been established, 
except for the particular cases,
such as  
theories with 
the canonical kinetic term~\cite{Bekenstein:1972ny,Bekenstein:1995un},
the noncanonical kinetic terms \cite{Graham:2014mda},
and 
the nonminimal coupling to the scalar curvature \cite{Sotiriou:2011dz,Faraoni:2017ock}.
To be more specific,
we focus on the Horndeski theory
described by the action~\eqref{action},
which satisfy the recent bound on the propagation speed of GWs,
derive the sufficient conditions
that allow the Ricci-flat metric solutions with the nontrivial profile for the scalar field,
and apply them to obtain the stealth Schwarzschild solutions 
in the shift-symmetry breaking theories.

The paper is organized as follows:
In Sec.~\ref{sec2}, we review the scalar-tensor theory \eqref{action}
and derive the equations of motion. 
In Sec.~\ref{sec3}, we covariantly derive the conditions 
for the theory \eqref{action} to allow general Ricci-flat solutions with 
the nontrivial profile of the scalar field.
In Sec.~\ref{sec4}, we focus on the Schwarzschild metric
and derive the stealth Schwarzschild solutions
for the ansatz where the scalar field is a function of the radial coordinate, $\phi=\phi_0(r)$.
We present two types of stealth Schwarzschild solutions, 
and show that, for the first solution, the scalar field is regular on the event horizon.
We then study the linear perturbation analysis about the solution, 
and clarify that
the kinetic term of the scalar perturbation in the second order action vanishes, 
indicating that the scalar mode is strongly coupled.
We also argue that stealth Schwarzschild solution with more general time-independent scalar field 
generically exhibits the same nature.
In Sec.~\ref{sec5}, we consider time dependent scalar field, and
also argue the nonexistence of stealth Schwarzschild solution. 
We explicitly show it for sum and product separable ansatze on the scalar field profile.
Finally, Sec.~\ref{sec6} is devoted to a brief summary and conclusion.

%%%%%%%%%%%%%%%%%%%%%%%%%%
\section{Setup}
\label{sec2}

\subsection{Equations of motion}
\label{sec21}

We consider the class of the Horndeski theory \eqref{action}.
Here, 
we do not assume the shift-symmetry in the scalar sector,
and in general $G_2$ and $G_3$ explicitly depend 
on the scalar field $\phi$ as well as $X$.
Note that 
in the theory \eqref{action}
the propagation speed of GWs
in the cosmological background
coincides with the speed of light.

Varying the action \eqref{action} with respect to the metric $g_{\mu\nu}$,
we obtain the gravitational equations of motion 
\begin{align}
\label{grav}
0
={\cal E}_{\mu\nu}
:=
&
-\frac{1}{2}G_{2X}\phi_\mu\phi_\nu
 -\frac{1}{2} G_2 g_{\mu\nu}
 +\frac{1}{2}G_{3X}\phi_\mu\phi_\nu\Box\phi 
 + \phi_{(\mu} \nabla_{\nu)} G_3 
 -\frac{1}{2}g_{\mu\nu} \phi_\lambda \nabla^\lambda G_3,
\nonumber\\
&+G_4 G_{\mu\nu} 
 +g_{\mu\nu} \left(G_{4\phi}\Box \phi -2X G_{4\phi\phi}\right)
 -G_{4\phi}\phi_{\mu\nu}
 -G_{4\phi\phi}\phi_\mu\phi_\nu,
\end{align}
where $G_{\mu\nu}$ is the Einstein tensor associated with respect to $g_{\mu\nu}$,
$G_{i\phi}$ and $G_{iX}$ are partial derivatives of $G_i(\phi,X)$
with respect to $\phi$ and $X$, respectively,
and $A_{(\mu\nu)}:=(A_{\mu\nu}+A_{\nu\mu})/2$.

Varying the action \eqref{action} with respect to the scalar field $\phi$,
we obtain the scalar field equations of motion 
\begin{align}
\label{sca}
0={\cal S}
:=
\nabla^\mu
\left(
 -G_{2X}\phi_\mu 
 +G_{3X}\phi_\mu\Box\phi 
+\nabla_\mu G_3
\right)
- \left(
  G_{2\phi}
- G_{3\phi}\Box\phi
+G_{4\phi}R  
\right).
\end{align}
These equations are not all independent, but constrained by 
\begin{align}
\label{constraint}
\nabla_\nu {\cal E}^\nu{}_\mu= \frac{\cal S}{2}\nabla_\mu \phi,
\end{align}
which is obtained by
the Bianchi identity.
Thus, the scalar field equation does not need to be considered separately.
As we shall see below,
once one obtains the conditions that ensure
the gravitational equations of motion ${\cal E}_{\mu\nu}=0$
to be satisfied,
one also obtains $\nabla_\nu {\cal E}^{\nu}{}_{\mu}=0$
if the conditions are conserved along the solution
(See Sec.~\ref{sec22}),
and then 
the scalar field equation of motion ${\cal S}=0$ is automatically satisfied
via Eq.~\eqref{constraint}.

\subsection{Ricci-flat solutions}
\label{sec22}

First, 
we consider the general spacetimes satisfying the Ricci-flat condition
\begin{align}
\label{ricci}
R_{\mu\nu}[g_{\alpha\beta}]=0.
\end{align}
They correspond to the vacuum solutions in GR
including the Schwarzschild and Kerr solutions under the certain symmetries.
In Sec.~\ref{sec3},
we will derive the conditions
on coupling functions
for the existence of the nontrivial profile of the scalar field $\phi=\phi_0(x^\mu)$
with 
\begin{align}
X_0=-\frac{1}{2}g^{\mu\nu}\phi_{0\mu}\phi_{0\nu}.
\end{align}
In the following, for any function $A=A(\phi,X)$,
$\partial_\mu A |_{\phi\to \phi_0,X\to X_0}$
represents that
\begin{align}
 \partial_\mu A\Big|_{\phi\to \phi_0,X\to X_0}
&:=
\left(A_{\phi}\,  \phi_{\mu}\right)\Big|_{\phi\to \phi_0, X\to X_0}
+
\left(A_{X} \, \partial_\mu X\right)\Big|_{\phi\to \phi_0, X\to X_0}
\nonumber\\
&=
A_{\phi} (\phi_0,X_0) \phi_{0\mu}
+A_{X} (\phi_0,X_0) \partial_\mu X_0,
\end{align}
where 
$A_\phi:=\partial A/\partial \phi$
and 
$A_X:=\partial A/\partial X$.
If the condition $A(\phi_0, X_0)={\rm const.}$
is satisfied on a trajectory $(\phi,X)=(\phi_0,X_0)$,
\begin{align}
\label{naba} 
\partial_\mu A\Big |_{\phi\to \phi_0,X\to X_0}=0,
\end{align}
also has to be satisfied as the consistency condition.

Moreover, 
we focus on the case of the minimal coupling of the scalar field
to gravity,
\begin{align}
\label{eh}
G_4=\frac{\Mpl^2}{2},
\end{align}
where $\Mpl^2:=(8\pi G)^{-1/2}$ is the reduced Planck mass squared
and $G$ is the gravitational constant.
Note that 
in this paper we set the speed of light and the Planck constant to unity,
i.e., $c=\hbar=1$.
As we will see later,
what is more important for obtaining a stealth Ricci-flat solution
is the existence of the nontrivial functions $G_2(\phi,X)$ and $G_3(\phi,X)$,
and no stealth Ricci-flat solution can be obtained 
only from the nontrivial $G_4(\phi)$.
Thus, 
the restriction to the case of Eq.~\eqref{eh} 
does not spoil the essence for the existence 
of a stealth Ricci-flat solution.
%%%%%%%%%%%

Let us remark on the conformal and disformal transformations.
Under the transformation
\be g_{\mu\nu} \to \alpha(\phi) g_{\mu\nu} + \beta(\phi)\pa_\mu\phi\pa_\nu\phi, \ee
with $\alpha=\alpha(\phi)$ and $\beta=\beta(\phi)$, 
the structure of the Horndeski Lagrangian does not change. 
Thus, one may expect that 
the theory~\eqref{action} with Eq.~\eqref{eh} 
may be conformally or disformally transformed to GR.
However, 
this is not the case,
as starting from the Einstein frame with Eq.~\eqref{eh}
the conformal transformation with $\alpha=\alpha(\phi)$ and $\beta=0$ 
yields $G_4=G_4(\phi)$,
and the disformal transformation with $\alpha=1$ and $\beta=\beta(\phi)$ 
yields $G_4=G_4(\phi,X)$ in the new frame.
Furthermore,
if a Ricci-flat solution exists in the original frame with Eq.~\eqref{eh},
the corresponding solution in the new frame
would be given by a BH hairy solution
with a non-Ricci-flat metric
and a nontrivial profile of the scalar field.

\subsection{Static and spherically symmetric spacetime}
\label{sec23}

After covariantly analyzing general conditions for the Ricci-flat solutions \eqref{ricci} in Sec.~\ref{sec3}, 
we shall focus on a static and spherically symmetric spacetime
\begin{align}
\label{sss}
g_{\mu\nu}dx^\mu dx^\nu
=-f(r)dt^2 +\frac{dr^2}{h(r)}+r^2 (d\theta^2+\sin^2\theta d\varphi^2),
\end{align}
where 
$t$, $r$, $a=(\theta,\varphi)$
are the time, radial and angular coordinates, 
respectively.
The $f$ and $h$ are the functions of $r$.
Because of the uniqueness theorem,
the static, spherically symmetric, 
and asymptotically flat solutions satisfying 
the vacuum Einstein equation \eqref{ricci}
is only the Schwarzschild solution
\begin{align}
\label{sch}
f=h=1-\frac{2M}{r}.
\end{align}
Below we shall derive conditions on the coupling functions 
for the existence of the nontrivial profile of the scalar field.
On the Schwarzschild background \eqref{sch},
we will focus on the following two cases
\begin{enumerate}
\item{The scalar field is solely the function of $r$, $\phi=\phi(r)$ (Sec.~\ref{sec4}),}
\item{The scalar field can also depend on the time and other spatial coordinates, $\phi=\phi(t,r,\theta,\varphi)$ (Sec.~\ref{sec5}).}
\end{enumerate}

%%%%%%%%%%%%%%%%%%%%%%%%%%
\section{Conditions for the stealth Ricci-flat solutions}
\label{sec3}

In this section we provide the covariant analysis.
Generalizing the strategy adopted in \cite{Motohashi:2018wdq} for a constant scalar field, 
we clarify conditions on $G_2(\phi,X)$ and $G_3(\phi,X)$ for the equations of motion
to allow general Ricci-flat solutions with nontrivial scalar field profile.
We stress that the analysis in this section applies 
general Ricci-flat solutions including 
Schwarzschild and Kerr solutions.
We shall show that breaking the shift symmetry is crucial for the following analysis, 
and that with shift symmetry one would not obtain nontrivial solution.

\subsection{The model with $G_3(\phi,X)=0$}
\label{sec31}

First, we consider the action~\eqref{action} with Eq.~\eqref{eh} and 
\begin{align}
\label{general_k}
G_2(\phi,X)\neq 0,
\qquad
G_3(\phi,X)=0,
\end{align}
which corresponds to
the Einstein gravity coupled to the k-essence type scalar field.

In the model \eqref{general_k},
if we impose the Ricci-flat condition \eqref{ricci} for the metric
and assume the existence of nontrivial profile of the scalar field $\phi=\phi_0(x^\mu)$,
the gravitational and scalar field equations of motion, i.e.,
\eqref{grav} and \eqref{sca} respectively,
reduce to 
\begin{align}
\label{grav_02}
0&={\cal E}_{\mu\nu}
=
-\frac{1}{2}G_{2X} \phi_{0\mu}\phi_{0\nu}
-\frac{1}{2} G_2 g_{\mu\nu},
\\
\label{sca02}
0&={\cal S}
=
-\phi_{0\mu} \nabla^\mu G_{2X}
-G_{2X} \Box \phi_{0}
-G_{2\phi}.
\end{align}
Note that $G_2$ and its derivatives in the right-hand sides are evaluated on $(\phi,X)=(\phi_0,X_0)$, and $\Box\phi_0:= \Box\phi|_{\phi\to \phi_0,X\to X_0}$.
Assuming the existence of the nontrivial scalar field $\phi=\phi_0(x^\mu)$
with $\phi_{0\mu}\neq 0$ and $X_0\neq 0$,
the trace of Eq.~\eqref{grav_02},
${\cal E}^\mu{}_\mu=0$,
gives
\begin{align}
\label{g2g2x}
G_{2X}(\phi_0,X_0)
=\frac{2G_2(\phi_0,X_0)}{X_0}.
\end{align}
Substituting it back into Eq.~\eqref{grav_02}, we obtain
\begin{align}
{\cal E}_{\mu\nu}
=-\frac{1}{2}G_2(\phi_0,X_0)
\left(
\frac{2\phi_{0\mu}\phi_{0\nu}}{X_0}
+g_{\mu\nu}
\right),
\end{align}
which yields the condition
\begin{align}
\label{g2cond1} 
&
G_2(\phi_0,X_0)=0.
\end{align}
Substituting Eq.~\eqref{g2cond1} into Eq.~\eqref{g2g2x}, we obtain
\begin{align}
\label{g2cond2}
&
G_{2X}(\phi_0,X_0)
=0.
\end{align}
As we mentioned in \eqref{naba}, for \eqref{g2cond1} to be satisfied on the trajectory $(\phi,X)=(\phi_0,X_0)$, 
the consistency condition $\partial_\mu G_2 |_{\phi\to \phi_0, X\to X_0}=0$ 
should be satisfied. Combining it with \eqref{g2cond2}, we obtain
\begin{align}
0=
\partial_\mu G_2
\Big|_{\phi\to \phi_0, X\to X_0}
=
G_{2\phi}(\phi_0,X_0)\phi_{0\mu}
+G_{2X} (\phi_0,X_0)\partial_\mu X_0
=G_{2\phi}(\phi_0,X_0)\phi_{0\mu}
=0,
\end{align}
which leads to 
\begin{align}
\label{g2cond3}
G_{2\phi}(\phi_0,X_0)=0.
\end{align}
Likewise, the consistency conditions for Eqs.~\eqref{g2cond1} and \eqref{g2cond2} are given by 
\begin{align}
\label{g2cond4}
&\partial_\mu G_{2\phi}
\Big|_{\phi\to \phi_0,X\to X_0}=0, \qquad 
\partial_\mu G_{2X}
\Big|_{\phi\to \phi_0,X\to X_0}=0.
\end{align}
With Eqs.~\eqref{g2cond2}, \eqref{g2cond3}, and \eqref{g2cond4},
the scalar field equation of motion \eqref{sca02} is satisfied.

It is worthwhile to note here that it is impossible to obtain nontrivial model that satisfies these requirements.  
Indeed, if one focuses on
the shift-symmetric model with $G_2=G_2(X)$, 
the condition~\eqref{g2cond4} reduce to 
$\partial_\mu G_{2X}(X_0)=0$. 
Considering its consistency condition with $\partial_\mu X_0\neq 0$, 
we obtain $G_{2XX}(X_0)=0$.
In such a way, we successively obtain
$G_{2X}(X_0)=G_{2XX}(X_0)=G_{2XXX}(X_0)=\cdots=0$.
Consequently, the only possible option is the trivial model $G_2(X)=0$.

On the contrary, in 
the model without the shift symmetry $G_2=G_2(\phi,X)$,
the condition~\eqref{g2cond4} generates 
\be \label{g2cond5}
\bem 
G_{2\phi\phi} & G_{2\phi X}\\
G_{2\phi X} & G_{2XX} 
\eem
\bem \phi_{0\mu} \\ \partial_\mu X_{0} \eem
= 0 , \ee
where the arguments of the matrix are evaluated at $(\phi,X)=(\phi_0,X_0)$.
In order for \eqref{g2cond5} to be compatible with nontrivial solution 
with $(\phi_{0\mu},\partial_\mu X_{0})\neq 0$, 
the necessary and sufficient condition is the degeneracy of the matrix, i.e.,
\begin{align}
\label{g2der}
G_{2\phi\phi}(\phi_0,X_0)
G_{2XX} (\phi_0,X_0)
-
G_{2X\phi}(\phi_0,X_0)^2
=0.
\end{align}

We assume that $G_2(\phi,X)$ is given by
\be G_2=f_2\left[ g_2(\phi,X)\right], \ee 
where $f_2(y)$ is a regular function of $y$, and $g_2(\phi,X)$ is a regular function of $(\phi,X)$.
The existence of stealth Ricci-flat solution on the trajectory $g_2(\phi_0, X_0)=c_2$,
where $c_2$ is a constant,
requires the conditions \eqref{g2cond1}, \eqref{g2cond2}, and \eqref{g2cond3},
which yield $f_2=f_2'=0$ at $g_2(\phi_0, X_0)=c_2$,
where Eq.~\eqref{g2der} is also satisfied.
Since $f_2$ is a regular function,
it can be written as a series expansion 
with respect to $(g_2(\phi,X)-c_2)$ 
consisting of terms of more than the second order:
\begin{align}
\label{g2gen}
G_2(\phi,X)= 
M_2^4
\sum_{n=2}^\infty
\gamma_{2,n}
\left[
g_2(\phi,X)
-c_2
\right]^n,
\end{align}
where
$\gamma_{2,n}$ ($n\geq 2$) is a dimensionless constant,
and $M_2$ is a constant of mass dimension one.
%%%%%
To be more specific,
we focus on the case where $g_2$ is a linear function of $\phi$ and $X$ and set $c_2=0$
without loss of generality,
\begin{align}
\label{g2conc}
G_2(\phi,X)= 
M_2^4
\sum_{n=2}^\infty
\gamma_{2,n}
\left(
\frac{m_2\phi}{M_2^2}
+
\frac{X}{M_2^4}
\right)^n,
\end{align}
where $m_2$ is a constant of mass dimension one.
In this model,  
the stealth scalar field satisfies
\begin{align}
\label{g2conc2}
X_0=-m_2 M_2^2 \phi_0.
\end{align}
It can be solved as a differential equation for $\phi_0(x^\mu)$ for a specific case.
We will provide an analytic solution with specific ansatz on the metric and scalar field in Sec.~\ref{sec4}.

\subsection{The model with $G_2(\phi,X)=0$}
\label{sec32}

Next, let us consider the action~\eqref{action} with \eqref{eh} and
\begin{align}
\label{general_gal}
G_2(\phi,X)=0,
\qquad
G_3(\phi,X)\neq 0,
\end{align}
which corresponds to the Einstein gravity coupled to the generalized galileon.

In the model \eqref{general_gal},
if we impose the Ricci-flat condition \eqref{ricci} for the metric
and assume the existence of nontrivial profile of the scalar field $\phi=\phi_0(x^\mu)$,
the gravitational and scalar field equations of motion,
\eqref{grav} and \eqref{sca},
respectively,
reduce to 
\begin{align}
\label{grav03}
&0
={\cal E}_{\mu\nu}
=
 \frac{1}{2} \phi_{0\mu}\phi_{0\nu} G_{3X}\Box\phi_0 
 + \phi_{0(\mu} \nabla_{\nu)} G_3 
 -\frac{1}{2}g_{\mu\nu} \phi_{0\lambda} \nabla^\lambda G_3 , \\
\label{sca03}
&0={\cal S}
=
\nabla^\mu
\left(
 \phi_{0\mu} G_{3X}\Box\phi_0 
+\nabla_\mu G_3
\right)
+ G_{3\phi}\Box\phi_0.
\end{align}
The trace of Eq.~\eqref{grav03}, ${\cal E}^\mu{}_\mu=0$,
gives the condition
\begin{align}
\label{g3cond1}
\phi_\lambda \nabla^\lambda G_{3}  \Big|_{\phi\to \phi_0,X\to X_0}
=-X_0
 G_{3X} (\phi_0,X_0)\Box\phi_0.
\end{align}
Plugging \eqref{g3cond1} back into Eq.~\eqref{grav03}, we obtain
\begin{align}
0= {\cal E}_{\mu\nu}
=-\frac{1}{2}
  \left(
   \frac{\phi_{0\mu} \phi_{0\nu}}{X_0}
+ g_{\mu\nu}
  \right) 
\phi_\lambda \nabla^\lambda G_3 \Big|_{\phi\to \phi_0,X\to X_0}
+\phi_{(\mu}\nabla_{\nu)} G_3 
\Big|_{\phi\to \phi_0,X\to X_0},
\end{align}
which is satisfied if
\begin{align}
\label{g3cond5}
\partial_\mu G_{3} \Big|_{\phi\to \phi_0, X\to X_0}=0.
\end{align}

Parallel to Sec.~\ref{sec31}, we can see that \eqref{g3cond5} implies that 
only models breaking the shift symmetry can 
generate a nontrivial solution.  
Indeed, for $G_3=G_3(X)$, the condition \eqref{g3cond5} reduces to $G_{3X}(X_0)=0$, 
and from the consistency condition with $\partial_\mu X_0\neq 0$,
one successively obtains 
$G_{3XX}(X_0)=G_{3XXX}(X_0)=\cdots=0$,
leaving the trivial model $G_3(X)=0$.
Hence, below we consider models breaking the shift symmetry: $G_3=G_3(\phi, X)$ with $G_{3\phi}\neq 0$.

Plugging \eqref{g3cond5} into \eqref{g3cond1}, so long as $X_0\neq 0$, we obtain 
\be \label{g3xbox} G_{3X}(\phi_0,X_0)\Box\phi_0=0. \ee
The consistency conditions for \eqref{g3cond5} and \eqref{g3xbox} yield 
$\Box G_3 = 0$ and $\partial_\mu(G_{3X}\Box\phi_0)=0$, 
which guarantees that the first term of the scalar field equation of motion~\eqref{sca03} vanishing,
and the remaining term is $G_{3\phi}\Box\phi_0$.
We can check that the remaining term is also vanishing for each branch of \eqref{g3xbox}:
\begin{align}
\label{g3cond2}
&{\rm 1)}\qquad \Box\phi_0=0, \\
\label{g3cond3}
&{\rm 2)}\qquad  G_{3X}(\phi_0,X_0)=0.
\end{align}

\subsubsection{Case 1}
\label{sec411}

Case 1 constrains the scalar field profile, 
and there are no further constraints for the functional form of $G_3$ rather than \eqref{g3cond5}. 
The scalar field profile is then determined by solving the differential equation~\eqref{g3cond2}
with specific ansatz on the metric and scalar field.

\subsubsection{Case 2}
\label{sec412}

In Case 2, 
from Eqs.~\eqref{g3cond5} and \eqref{g3cond3}, we obtain
\begin{align}
\label{g3cond4}
G_{3\phi}(\phi_0,X_0)=0,
\end{align}
which guarantees the remaining term $G_{3\phi}\Box\phi_0$ 
in the scalar field equation of motion~\eqref{sca03} vanishing.
The conditions \eqref{g3cond5}, \eqref{g3cond3}, and \eqref{g3cond4}
are sufficient to have the stealth Ricci-flat solutions with $\phi_{0\mu}\neq 0$ and $\partial_\mu X_0\neq 0$.

The following process is then parallel to Sec.~\ref{sec31}.
The consistency conditions of \eqref{g3cond3} and \eqref{g3cond4} provide
\be \label{g3cond}
\bem 
G_{3\phi\phi} & G_{3\phi X}\\
G_{3\phi X} & G_{3XX} 
\eem
\bem \phi_{0\mu} \\ \partial_\mu X_{0} \eem
= 0,
\ee
which implies that nontrivial solution with $(\phi_{0\mu},\partial_\mu X_{0})\neq 0$ 
exists if and only if the condition 
\begin{align}
\label{g3der2}
G_{3\phi\phi}(\phi_0,X_0)
G_{3XX} (\phi_0,X_0)
-
G_{3X\phi}(\phi_0,X_0)^2
=0,
\end{align}
is satisfied.
As for the case \eqref{general_k},
a general  model satisfying 
Eqs.~\eqref{g3cond3}, \eqref{g3cond4}, and \eqref{g3der2}
is given by 
\begin{align}
\label{g3gen}
G_3(\phi,X)= 
M_3
\sum_{n=2}^\infty
\gamma_{3,n}
\left[
g_3(\phi,X)
-c_3
\right]^n,
\end{align}
where $M_3$ is a constant of mass dimension one.
The stealth Ricci-flat solution exists for the trajectory $g_3(\phi_0, X_0)=c_3$,
where $c_3$ is a constant.
%%%%%
We focus on the case where $g_3$ is a linear function of $\phi$ and $X$ and set $c_3=0$,
\begin{align}
\label{g3conc}
G_3(\phi,X)= 
M_3
\sum_{n=2}^\infty
\gamma_{3,n}
\left(
\frac{m_3\phi}{M_3^2}
+
\frac{X}{M_3^4}
\right)^n,
\end{align}
where $\gamma_{3,n}$ ($n\geq 2$) is a dimensionless constant,
and $m_3$ is a constant of mass dimension one.
In this model,
the stealth scalar field satisfies
\begin{align}
\label{g3conc2}
X_0=-m_3 M_3^2 \phi_0. 
\end{align}

\subsection{The model with $G_2(\phi,X)\neq 0$ and $G_3(\phi,X)\neq 0$}
\label{sec33}

Finally, we consider the model with 
\begin{align}
\label{general_gg}
G_2(\phi,X)\neq 0,
\qquad
G_3(\phi,X)\neq 0.
\end{align}

If we impose the Ricci-flat condition \eqref{ricci} for the metric
and assume the existence of nontrivial profile of the scalar field $\phi=\phi_0(x^\mu)$,
the gravitational and scalar field equations of motion,
\eqref{grav} and \eqref{sca},
respectively,
reduce to 
\begin{align}
\label{grav2}
0
&={\cal E}_{\mu\nu}
=
-\frac{1}{2}\phi_{0\mu}\phi_{0\nu}G_{2X}
 -\frac{1}{2} G_2 g_{\mu\nu}
 +\frac{1}{2} \phi_{0\mu}\phi_{0\nu} G_{3X}\Box\phi_0
 + \phi_{0(\mu} \nabla_{\nu)} G_3 
 -\frac{1}{2}g_{\mu\nu} \phi_{0\lambda} \nabla^\lambda G_3 ,
\\
\label{sca2}
0
&={\cal S}
=
\nabla^\mu
\left(
 -\phi_{0\mu}G_{2X} 
 +\phi_{0\mu}G_{3X}\Box\phi_0 
 +\nabla_\mu G_3
\right)
- G_{2\phi}
+ G_{3\phi}\Box \phi_{0},
\end{align}
respectively.

The trace of Eq.~\eqref{grav2}, 
${\cal E}^\mu{}_\mu=0$,
is given by 
\begin{align}
\label{mix0}
\left(
G_{2X}(\phi_0,X_0)
-G_{3X}(\phi_0,X_0)
\Box\phi_0
\right)X_0
=
2G_2 (\phi_0,X_0)
+\phi_{\lambda}\nabla^\lambda G_3 
 \Big|_{\phi\to\phi_0,X\to X_0}.
\end{align}
Substituting Eq.~\eqref{mix0} into Eq.~\eqref{grav2}, we obtain
\begin{align}
\label{traceless}
0=
{\cal E}_{\mu\nu}
&=
-
\left(
G_2+\phi_\lambda\nabla^\lambda G_3
\right)
\left(
\frac{\phi_{0\mu}\phi_{0\nu}}{X_0}
+\frac{1}{2}g_{\mu\nu}
\right)
+\frac{G_{3X}}{2}
\left(
\frac{\phi_{0\mu} \phi_{0\nu}\phi_{0\lambda}\nabla^\lambda X_0 }{X_0}
+2\phi_{0(\mu}\nabla_{\nu)} X_{0}
\right),
\end{align}
which is satisfied if 
\begin{align}
\label{mix1}
&G_2(\phi_0,X_0) 
+\phi_{\lambda}\nabla^\lambda G_3
 \Big|_{\phi\to\phi_0,X\to X_0}
=0, \\
\label{mix12}
&G_{3X}(\phi_0,X_0)=0.
\end{align}
From Eqs.~\eqref{mix1} and \eqref{mix12},
\begin{align}
 G_2 (\phi_0,X_0)
-2X_0G_{3\phi} (\phi_0,X_0) 
=0.
\label{mix22}
\end{align}
From Eqs.~\eqref{mix0}, \eqref{mix1}, and \eqref{mix12},
then
\begin{align}
G_{2X}(\phi_0,X_0)
-2G_{3\phi}(\phi_0,X_0) 
=0.
\label{mix32}
\end{align}
The consistency condition for Eqs.~\eqref{mix12}--\eqref{mix32} yield
\begin{align}
\label{mix4}
\partial_\mu G_{3X}\Big|_{\phi\to \phi_0,X\to X_0}=0,
\nonumber
\\
\partial_\mu (G_{2X}-2G_{3\phi})|_{\phi\to \phi_0,X\to X_0}=0,
\nonumber
\\
\partial_\mu(G_2-X G_{2X})|_{\phi\to \phi_0,X\to X_0}
=0.
\end{align}
Note that we arranged the third equation for a later convenience.
With Eqs.~\eqref{mix12},
\eqref{mix22}, 
\eqref{mix32},
and 
\eqref{mix4},
the scalar field equation of motion \eqref{sca2} is also satisfied.

As for the previous models,
a general model satisfying the requirement is then given by 
\begin{align}
\label{g2g3gen}
&G_2(\phi,X)
=
M_0^4
\sum_{n=2}^\infty
\gamma_{2,n}
\left[
g_0(\phi,X)-c_0
\right]^n,
\nonumber
\\
&
G_3(\phi,X)
=
M_0
\sum_{n=2}^\infty
\gamma_{3,n}
\left[
g_0(\phi,X)-c_0
\right]^n,
\end{align}
where 
$\gamma_{2,n}$ and $\gamma_{3,n}$ are dimensionless constants,
$M_0$ is a constant of mass dimension one,
and the solution $g_0(\phi_0,X_0)=c_0$ gives the stealth Ricci-flat solution.
We focus on the case where $g_0$ is a linear function of $\phi$ and $X$ and set $c_0=0$,
\begin{align}
\label{g2g3conc}
&G_2(\phi,X)
=
M_0^4
\sum_{n=2}^\infty
\gamma_{2,n}
\left(
\frac{m_0\phi}{M_0^2}
+
\frac{X}{M_0^4}
\right)^n,
\nonumber
\\
&
G_3(\phi,X)
=
M_0
\sum_{n=2}^\infty
\gamma_{3,n}
\left(
\frac{m_0\phi}{M_0^2}
+
\frac{X}{M_0^4}
\right)^n,
\end{align}
where 
$\gamma_{2,n}$ and $\gamma_{3,n}$ are dimensionless constants,
and $m_0$ is constant of mass dimension one.
In the model,
the stealth scalar field in the model \eqref{g2g3conc} obeys
\begin{align}
\label{g2g3traj}
X_0=-m_0 M_0^2\phi_0.
\end{align}

%%%%%%%%%%%%%%%%%%%%%%%%%%%%%%%%%%%%%%%%%%%%%%%%%%%%
\section{Stealth Schwarzschild solution with $\phi_0=\phi_0(r)$}
\label{sec4}

In Secs.~\ref{sec4} and \ref{sec5}, 
we assume that the metric is given by the Schwarzschild solution~\eqref{sss} with \eqref{sch}
and derive the conditions for the existence of
a nontrivial profile of the scalar field. 
In this section,
we adopt the ansatz $\phi=\phi_0(r)$, where $X_0=-(h/2)\phi_0'(r)^2$.
We shall derive a nontrivial solution of $\phi=\phi_0(r)$, which is unique for theory breaking the shift symmetry.

\subsection{The model with $G_3(\phi,X)=0$}
\label{sec41}

First, we focus on the model \eqref{general_k}.
A concrete model that satisfies all the conditions is given by Eq.~\eqref{g2conc},
for which the stealth scalar field  
satisfies Eq.~\eqref{g2conc2}.
For the Schwarzschild spacetime with the ansatz $\phi=\phi_0(r)$, Eq.~\eqref{g2conc2} reads
\be 
\phi_0'^2=\f{2m_2M_2^2}{1-2M/r}\phi_0.
\ee
The solution of the scalar field is then found to be
\be \label{sfc}
\phi_0(r) = 2 m_2M_2^2 M^2 \kk{ \sqrt{x} \sqrt{x - 1 } + \ln\mk{ \sqrt{x} + \sqrt{x - 1 } } - C_2}^2 ,\ee
where $x:=r/(2M)$, and
$C_2$ is a dimensionless integration constant.
At the vicinity of the event horizon $r= 2M$, 
the scalar field behaves as
\begin{align} 
\f{\phi_0(r)}{2 m_2M_2^2 M^2} &= \kk{ \sqrt{x} \sqrt{x - 1 } + \ln\mk{ \sqrt{x} + \sqrt{x - 1 } } }^2  - 2C_2 \kk{ \sqrt{x} \sqrt{x - 1 } + \ln\mk{ \sqrt{x} + \sqrt{x - 1 } } } + C_2^2  \notag\\
&= 4 (x - 1) + \f{4}{3} (x - 1)^2 - \f{4}{45} (x - 1)^3 + \cdots +C_2\sqrt{x - 1}\kk{ -4  - \f{2}{3} (x - 1) + \f{1}{10} (x - 1)^2 - \cdots } + C_2^2 .
\end{align}
We see that for $C_2=0$ all the terms with the half-integer powers of $(x-1)$ vanish,
and hence $\phi_0'(r), \phi_0''(r),\cdots$ are regular on the event horizon.

The solution~\eqref{sfc} is different from the stealth Schwarzschild solution~\cite{Babichev:2013cya,Kobayashi:2014eva} 
in shift symmetric theories from several aspects.
In the case of the stealth Schwarzschild solution in the shift symmetric Horndeski theories, 
the crucial point is that the scalar field has a different coordinate dependence than the metric, 
i.e., while the metric is Schwarzschild spacetime depending only on $r$, 
the scalar field has the linear time dependence as $\phi_0=qt+\psi(r)$ with $q=\text{const}$. 
This is compatible with the equations of motion since
the time dependence does not show up 
in the equations of motion, which is a natural consequence 
of the Lagrangian depending only on $r$ with the shift symmetry.
Consequently, only the parameter $q$ enters the equations of motion and one can derive analytic solution for $\psi(r)= q F(r)+\text{const}$, which is constant for the limit $q\to 0$.
Therefore, the shift symmetry and the linear time dependence 
play a crucial role to support the stealth Schwarzschild solution
in shift symmetric theories.

On the other hand, the solution~\eqref{sfc} only exists for theories breaking the shift symmetry
as we clarified in Sec.~\ref{sec3},
and the scalar field does share the same symmetry with the metric
(See also Sec.~\ref{sec5} for the case $\phi_0=\phi_0(t,r)$).
The effect of the nontrivial scalar profile to the metric sector is hidden in a nontrivial way
and the metric remains the Schwarzschild solution at the background level. 
Thus, the difference from GR would show up only at the level of perturbations.
We shall address the linear perturbation analysis in Sec.~\ref{sec44}.

\subsection{The model with $G_2(\phi,X)=0$}
\label{sec42}

Second, we focus on the model \eqref{general_gal}.
Eq.~\eqref{g3cond5} provides the condition
\begin{align}
\label{g3s1}
\partial_r G_{3} 
\Big|_{\phi\to \phi_0,X\to X_0}
=G_{3\phi} (\phi_0,X_0)\phi_0'
+G_{3X}(\phi_0,X_0)X_0'
=0,
\end{align}
and hence 
\begin{align}
G_3(\phi_0,X_0)={\rm const.}
\end{align}
Then, 
Eq.~\eqref{g3cond1} reduces to 
$G_{3X} (\phi_0,X_0)\Box\phi_0 =0$
which leads to  
\begin{align} 
\left[
r(r-2M)\phi_0''
+2(r-M)\phi_0'
\right]
G_{3X}(\phi_0,X_0)=0,
\end{align}
which provides Case 1 and Case 2 as discussed in Sec.~\ref{sec32}.

\subsubsection{Case 1}
\label{sec421}

In Case 1,
\begin{align}
r(r-2M)\phi_0''
+2(r-M)\phi_0'
=0.
\end{align}
The solution of $\phi_0$ is given by 
\begin{align}
\label{sol0}
\phi_0(r)=P+Q\ln \left(1-\frac{2M}{r}\right),
\end{align}
where $P$ and $Q$ are integration constants.
In this case, $G_3(\phi,X)$ is not be specified
except that it satisfies \eqref{g3s1}.
Unlike the solution~\eqref{sfc}, the solution~\eqref{sol0} is not regular at the event horizon,
unless $Q=0$ where the scalar field is trivial.

\subsubsection{Case 2}
\label{sec422}

In Case 2,
a concrete model is given by Eq.~\eqref{g3conc}.
The solution for the scalar field $\phi_0(r)$
is given by the same solution as Eq.~\eqref{sfc}
with the replacement $(M_2,m_2,C_2)\to(M_3,m_3,C_3)$, 
where $C_3$ is an integration constant, and the regularity of $\phi_0'(r)$ on the event horizon 
requires $C_3=0$.

\subsection{The model with $G_2\neq 0$ and $G_3\neq 0$}
\label{sec43}

Finally, 
we consider the model \eqref{general_gg}.
In the case of $\phi_0=\phi_0(r)$, 
Eq.~\eqref{mix0} and \eqref{mix1} have to be imposed.
However,
since the combination inside the round bracket of 
the second term in the right-hand side of \eqref{traceless}
trivially vanishes for all the components,
in general
$G_{3X}(\phi_0,X_0)=0$ 
does not need to be imposed.

If Eq.~\eqref{mix12} is imposed by hand,
since Eqs.~\eqref{mix22} and \eqref{mix32} are also satisfied, 
a concrete model is given by Eq.~\eqref{g2g3conc}.
The scalar field $\phi_0$ is given by Eq.~\eqref{sfc}
with the replacement of $(M_2,m_2,C_2)\to(M_0,m_0,C_0)$,
where $C_0$ is an integration constant, and the regularity 
of $\phi_0'(r)$, $\phi_0''(r)$, $\cdots$ on the event horizon requires $C_0=0$.

\subsection{Linear perturbations and the absence of the kinetic term}
\label{sec44}

Before closing this section,
let us mention the linear perturbations about the stealth Schwarzschild
solutions~\eqref{sfc} and \eqref{sol0},
and their stability. 
In Refs.~\cite{Kobayashi:2012kh,Kobayashi:2014wsa},
the odd- and even-parity perturbation analyses 
about general static and spherically symmetric BH solutions 
including Schwarzschild solution in the full Horndeski theory 
with the scalar field profile $\phi=\phi(r)$ were formulated, 
and  
the conditions for the stability and the propagation speeds were derived.
For the odd-parity mode, 
the conditions to evade ghost and gradient instabilities are
\be \label{odd-sta} \F>0,\quad \G>0, \quad \H>0, \ee
and the sound speed is given by
\be \label{odd-cs} c^2_{\rm odd} = \f{\G}{\F}, \ee
where $\F,\G,\H$ are defined by Eqs.~(17)--(19) in Ref.~\cite{Kobayashi:2012kh}. 
The odd mode is nonvanishing only for $\ell\geq 2$.
For the even-parity modes, the no-ghost condition is given by
\be \label{even-sta} \ell(\ell+1) \P_1 -\F >0 ,\quad 2\P_1 -\F > 0, \ee
where $\ell\geq 2$, and the sound speeds are given by
\be \label{even-cs} 
c^2_{{\rm even},1} = \f{\G}{\F} ,\quad 
c^2_{{\rm even},2}
=\f{(2r^2\Gamma\H
-\G\Xi)\Xi\phi'^2-4r^4\Sigma\H^2/h}{(2r\H+\Xi\phi')^2(2\P_1 -\F)}, \ee
where 
\be \label{eqP1} \P_1
= \f{1}{2} \f{r^2  \H^2}{2 r \H + \Xi \phi'} \f{d}{dr} \mk{ \ln \f{f}{h} } 
+ \f{d}{dr} \mk{ \f{r^2  \H^2}{2 r \H + \Xi \phi'}} , \ee
and $\Xi,\Gamma,\Sigma$ are defined by Eqs.~(36), (42), and (45) in Ref.~\cite{Kobayashi:2014wsa},
respectively.
The first even-parity mode is nonvanishing only for $\ell\geq 2$, whereas the second even-parity mode exists for all $\ell$ unless $2\P_1-\F=0$.
Note also that the first condition of \eqref{even-sta} for $\ell\geq 2$ is always satisfied if the second condition of \eqref{even-sta} is satisfied, while the opposite is not the case.  
The numerator of $c^2_{{\rm even},2}$ also needs to be positive to evade gradient instability.
Among the one odd-parity mode and two even-parity modes, 
the odd-parity mode and the first even-parity mode for $\ell\geq 2$
correspond to the tensor perturbations with respect to the three-dimensional space, 
i.e., they describe GWs.
On the other hand, the second even-parity mode,
which exists for all the $\ell$ modes unless $2\P_1-\F=0$,
corresponds to the scalar perturbation.
This mode highlights the deviation from GR most crucially.
For the Schwarzschild solution~\eqref{sch}, the first term of \eqref{eqP1} 
vanishes and we only need to look at $\F,\H,\Xi,\phi'$ to evaluate $2\P_1-\F$.
The class of the Horndeski theory 
in which $G_{3X}(\phi_0,X_0)=0$ where $(\phi_0,X_0)$ denotes the solution for the scalar field
(See Eqs.~\eqref{g3cond3} and \eqref{mix12})   
and $G_4$ and $G_5$ are constant 
yields $2\P_1-\F=0$ about the Schwarzschild background
(See Sec. \ref{sec441}). 
On the other hand, 
the second even-parity mode would be
propagating on the Schwarzschild background 
in the Horndeski theory other than this class.
Thus, the kinetic coupling 
of the scalar field to the spacetime curvature
due to the nontrivial $X$-dependent $G_4(\phi,X)$ and $G_5(\phi,X)$ is
crucial for the second even-parity mode on the Schwarzschild background to propagate.

For the theory~\eqref{action}, the functions are given by 
\begin{align} \label{FGH}
&\F=\G=\H=M_{\rm Pl}^2,
\quad \Gamma=-4XG_{3X},\quad \Xi=-2r^2XG_{3X}, \notag\\
&\Sigma= X\kk{G_{2X}+2XG_{2XX} - h\phi' \mk{\f{4}{r}+\f{f'}{f}} (XG_{3X})_X - 2(G_{3\phi}+ 2XG_{3\phi X}) } ,
\end{align}
where the functions in the right-hand sides are evaluated at $(\phi,X)=(\phi_0,X_0)$ 
and hence take different values for each stealth Schwarzschild solution.
First, let us focus on the GW modes, namely, the odd-parity mode and the first even-parity mode.
From \eqref{odd-cs}, \eqref{even-cs}, and \eqref{FGH}, 
it is clear that in the theory~\eqref{action} 
GWs propagate with the speed of light, 
i.e., $c^2_{\rm odd}=c^2_{{\rm even},1}=1$, the same as those in GR,
which satisfies the stringent observational constraint.
On the other hand, the scalar perturbation, namely,
the second even-parity mode
behaves differently for each stealth Schwarzschild solution.
As for the stability conditions, 
while the condition~\eqref{odd-sta} for the odd mode 
is always satisfied for the theory~\eqref{action}, 
the condition~\eqref{even-sta} and $c^2_{{\rm even},2}>0$ for the even-parity modes  
need to be studied separately for each stealth Schwarzschild solution.

We emphasize 
that 
the argument for obtaining the stealth Ricci-flat solution
in Sec.~\ref{sec3}
does not apply 
to the Horndeski theory with 
nontrivial $G_4(\phi,X)$ and $G_5(\phi,X)$,
since 
due to the kinetic coupling of the scalar field
to the spacetime curvature
the gravitational equations of motion
also  
depend on the Riemann tensor
and hence 
the conditions for the stealth Ricci-flat solution
cannot be specified only within the scalar field sector.
However,
as argued in Sec.~\ref{sec1},
the kinetic coupling in
the Horndeski theory 
with nontrivial $X$-dependent $G_4(\phi,X)$ and $G_5(\phi,X)$
would modify the propagation speed of gravitational waves
on cosmological backgrounds
which was significantly constrained by the latest measurements 
of a binary neutron star merger
and its associated short gamma-ray burst.
Also, by using the conformal transformation, theories with nontrivial $G_4(\phi)$ can be 
recast into the Einstein frame action.  
Thus, our analysis applies to the Horndeski subclass~\eqref{action}, and
the analysis of the Horndeski theory 
with nontrivial $X$-dependent $G_4(\phi,X)$ and $G_5(\phi,X)$
will have to be done separately.

We also emphasize
that even if the kinetic term of the second even-parity mode
is generically nonvanishing 
in a class of the Horndeski theory,
it might vanish at some radius.
Then, the strong coupling problem arises again.
Of course,
since this would depend on the choice of the
coupling functions in the Horndeski theory and background solution,
we also leave the further analysis for our future work.

\subsubsection{
Solution \eqref{sfc}}
\label{sec441}

First, let us check the perturbations about the first stealth Schwarzschild solution~\eqref{sfc}.
For this background, we obtain $\Gamma=\Xi=0$ and $\P_1=1/2$ 
as $G_{3X}(\phi_0,X_0)=0$ for the cases considered in Secs.~\ref{sec42}, \ref{sec43}, 
or $G_3$ itself vanishes entirely for the case considered in Sec.~\ref{sec41}.
We see that the first condition of \eqref{even-sta} is satisfied, 
whereas the second condition is not, as $2{\cal P}_1-{\cal F}=0$.
Thus, 
the kinetic term of the scalar perturbation in the second order action vanishes, 
indicating that the scalar mode is strongly coupled in
the stealth Schwarzschild solution~\eqref{sfc}. 
For such a solution,
higher order corrections are inevitably significant,
and the linear perturbation theory loses the predictability.

One might think that the absence of the kinetic term of the scalar mode in the quadratic action
might be avoidable,
if one introduces a new time coordinate of the Schwarzschild solution,
for instance,
the Eddington$-$Finkelstein 
or
Gullstrand$-$Painlev\'e coordinates \cite{Martel:2000rn},
where the metric tensor contains an off-diagonal component of the time and space,
and hence the Lagrangian 
would contain the nonzero kinetic term for the linear perturbations.
Moreover, in such a coordinate system,
the solution for the scalar field~\eqref{sfc} with $C_2=0$
can be analytically extended to the region inside the (future) event horizon.
However,
the characteristic curves of this solution
coincide with
the $t={\rm const.}$ surfaces originated from the bifurcation point
of the Schwarzschild spacetime,
which are spacelike outside the horizon
and timelike inside the horizon.
Hence, we cannot find a Cauchy surface
which can intersect all the characteristic curves,
and the initial value problem is still ill-defined at the liner perturbation level.
Indeed, the ill-posedness of the initial value problem is diffeomorphism invariant.
This situation is analogous 
to the case of a specific self-accelerating solution
in nonlinear massive gravity \cite{Motloch:2015gta}.
In such a solution,
even 
though the absence of the kinetic term in the quadratic action
may be avoided 
by an alternative choice of the time coordinate,
the linear perturbation analysis about the solution still loses the predictability.

One might also think that the absence of the kinetic term of the scalar mode in the quadratic action
arises because of the specific choice of  
the models \eqref{g2conc}, \eqref{g3conc}, and \eqref{g2g3conc}. 
However, we expect that it
also arises in any model Eq.~\eqref{general_k}
satisfying the conditions Eqs.~\eqref{g2cond1}, \eqref{g2cond2} , \eqref{g2cond3}, and \eqref{g2der}.
Considering a small perturbation in the scalar field sector,
$\phi=\phi_0(r)+\delta \phi$, 
and neglecting the perturbation of the metric,
since the time derivative term in $X$
becomes $X\supset  X_0+\dot{\delta \phi}^2/(2f)$,
where a `dot' denotes the derivative with respect to the time $t$,
the leading order kinetic term is given by 
\begin{align}
G_2(\phi_0+\delta \phi, X_0+\delta X)
&=
\frac{1}{2G_{2\phi\phi}}
\left(
 G_{2\phi\phi}\delta\phi
+ G_{2X\phi}\delta X
\right)^2
+{\cal O}
\left(\delta\phi^3,~\delta \phi^2\, \delta X,~\delta \phi\, \delta X^2,~\delta X^3\right)
\supset
\frac{G_{2X\phi}}{2f}
\delta\phi \, \dot{\delta\phi}^2.
\end{align}
Hence,
the kinetic term for the linear perturbation vanishes at the quadratic level
for more general model with $\phi_0=\phi_0(r)$.
Similar arguments also apply to
the other models \eqref{general_gal} and \eqref{general_gg}.
Furthermore,
following the same argument,
we expect that  
the kinetic term for the scalar mode vanishes at the quadratic level
in stealth Schwarzschild solution 
with more general time independent profile of the stealth scalar field, 
$\phi_0=\phi_0(r,\theta,\varphi)$.
Thus, the absence of the kinetic term of the scalar mode in the quadratic action
would be generic feature for stealth Schwarzschild solution with any time independent scalar field.
The possibility of stealth Schwarzschild solution 
with time dependent profile of the scalar field
will be discussed in Sec.~\ref{sec5}.

\subsubsection{Solution \eqref{sol0}}
\label{sec443}

Finally, we consider the second stealth Schwarzschild solution~\eqref{sol0}.
For this background, since $G_{3X}(\phi_0,X_0)\neq 0$, 
in general we have $2{\cal P}_1-{\cal F}\neq 0$ and $c^2_{{\rm even},2}$ does not blow up.
However, since the concrete form of $G_3(\phi,X)$ cannot be uniquely specified 
from our conditions,
we cannot determine whether the condition~\eqref{even-sta} and $c^2_{{\rm even},2}>0$ are satisfied.
Thus, for the second stealth Schwarzschild solution~\eqref{sol0},
the stability about the even-parity modes 
depends on the specific form of $G_3(\phi,X)$.

More specifically,
the nonzero kinetic term of the second even-parity mode
for the solution \eqref{sol0} arises, 
since the term $\Xi \phi'$
in the denominator of the second term in the right-hand side of Eq.~\eqref{eqP1}
is generically nonvanishing.
In the large distance region $r\gg 2M$,
where the metric approaches that of the flat spacetime,
from Eq.~\eqref{sol0}
we find that the leading order behavior is 
$\phi_0\sim P-2MQ/r$
and 
$X_0\sim -\phi_0'^2/2\sim 1/r^4$,
and
consequently
$\Xi\phi_0'/(2r{\cal H)}
=-rG_{3X}X_0\phi_0'/M_{\rm Pl}^2
\sim  G_{3X} /r^5$.
Hence,
at least for the models with 
% the regular 
$G_{3X}$
which is not growing faster than $r^5$,
in the large distance region $r\gg 2M$,
$\Xi\phi_0'$ becomes negligible to $2r{\cal H}$
and from Eqs.~\eqref{eqP1} and \eqref{FGH},
$2{\cal P}_1- {\cal F}$ approaches $0$.
Thus, the kinetic term of the second even-parity mode
vanishes asymptotically at the spatial infinity.
This is consistent with our intuition
that  
the kinetic term of the scalar field perturbation should vanish
in the flat spacetime
where the background scalar field vanishes,
since the scalar and metric perturbations should be decoupled
in the flat spacetime.

%%%%%%%%%%%%%%%%%%%%%%%%%%%%%%
\section{Stealth Schwarzschild solution with time dependent scalar field}
\label{sec5}

In this section,
we argue the nonexistence of stealth Schwarzschild solution
with the time dependent scalar field. 
For the most general ansatz of the stealth scalar field, $\phi=\phi_0(t,r,\theta,\varphi)$, for which
\begin{align}
X_0=\frac{1}{2}
\left(
      \frac{1}{f}\dot{\phi}_0^2
      -f \phi_0'^2
      -\frac{1}{r^2}\phi_{0,\theta}^2
      -\frac{1}{r^2\sin^2\theta} \phi_{0,\varphi}^2
\right).
\end{align}
We mainly focus on the model \eqref{general_k},
but the same conclusion holds
for the models \eqref{general_gal} and \eqref{general_gg}. 
For the model \eqref{general_k},
we impose the conditions \eqref{g2cond1}, \eqref{g2cond2}, and \eqref{g2cond3},
and in addition the conditions \eqref{g2cond4} reduce to
\begin{align}
\label{g2cond44}
\frac{G_{2\phi\phi}(\phi_0,X_0)}{G_{2X\phi}(\phi_0,X_0)}
=\frac{G_{2X\phi}(\phi_0,X_0)}{G_{2XX}(\phi_0,X_0)}
=-\frac{\dot{X}_0}{\dot{\phi}_0}
=-\frac{X_0'}{\phi_0'}
=-\frac{\partial_\theta X_0}{\phi_{0,\theta}}
=-\frac{\partial_\varphi X_0}{\phi_{0,\varphi}}.
\end{align}
Moreover, in the model \eqref{g2gen},
the stealth scalar field satisfies the equation $g_2(\phi_0,X_0)=c_2$,
and from Eq.~\eqref{g2cond44} we obtain 
\begin{align}
\label{t_eq}
\dot{X}_0&= -\frac{g_{2\phi}(\phi_0,X_0)}{g_{2X} (\phi_0,X_0)}  \dot{\phi}_0,
\\
\label{r_eq}
X_0'&= -\frac{g_{2\phi} (\phi_0,X_0)}{g_{2X} (\phi_0,X_0)} \phi_0',
\\
\label{th_eq}
\partial_\theta X_0 &= -\frac{g_{2\phi} (\phi_0,X_0)}{g_{2X}(\phi_0,X_0)}  \phi_{0,\theta},
\\
\label{vp_eq}
\partial_\varphi X_0&= -\frac{g_{2\phi} (\phi_0,X_0)}{g_{2X} (\phi_0,X_0)} \phi_{0,\varphi}.
\end{align}
Thus, a single scalar variable $\phi_0=\phi_0(t,r,\theta,\varphi)$
has to satisfy 
the four independent conditions Eqs.~\eqref{t_eq}--\eqref{vp_eq},
which already indicates 
that in general
there is no consistent solution for the stealth scalar field,
except for the no-hair Schwarzschild solution $\phi_0=0$.

%%%%%%%
In the rest of this section, focusing further on the specific model \eqref{g2conc},
we will present the cases
for which we can explicitly see the nonexistence of stealth Schwarzschild solution.

\subsection{Case $\phi_0=\phi_0(t,\theta,\varphi)$}
\label{sec51}

Among the time dependent profile of the stealth scalar field,
the obvious example for the nonexistence of stealth solution is given by
$\phi_0=\phi_0(t,\theta,\varphi)$.
In this case, Eq.~\eqref{g2conc2} reduces to 
\begin{align}
\label{g2conc2_ttp}
\frac{1}{f(r)}\dot{\phi}_0^2
      -\frac{1}{r^2}\phi_{0,\theta}^2
      -\frac{1}{r^2\sin^2\theta} \phi_{0,\varphi}^2
=-2m_2 M_2^2\phi_0.
\end{align}
Even if Eq.~\eqref{g2conc2_ttp} is satisfied for a particular value of $r(>2M)$,
it fails to be satisfied for a different value of $r$.
Thus, there is no solution of the stealth scalar field $\phi_0$
for any $r$, except for the no-hair solution $\phi_0=0$.
The similar argument
applies to the other particular models \eqref{g3conc} and \eqref{g2g3conc}.
The same conclusion can be deduced 
for the restricted assumptions,
$\phi=\phi_0(t)$, $\phi_0(t,\theta)$, and $\phi_0(t,\varphi)$.

For the other profiles of the stealth scalar field,
$\phi=\phi_0(t,r)$,
$\phi_0(t,r,\theta)$,
$\phi_0(t,r,\varphi)$,
and 
$\phi_0(t,r,\theta,\varphi)$,
we need the further restrictions 
for the dependence on the variables.

\subsection{Case $\phi_0=\phi_0(t,r)$}
\label{sec52}

Next, we consider the scalar field profile $\phi_0=\phi_0(t,r)$.
If we focus on the model \eqref{g2conc},
only Eqs.~\eqref{t_eq} and \eqref{r_eq} are nontrivial.

\subsubsection{Case $\phi_0(t,r)= \chi(t)+\psi(r)$} 
\label{sec521}

First, we assume the sum separable ansatz for the scalar field $\phi_0=\chi(t)+\psi(r)$,
for which Eq.~\eqref{t_eq} reads
\be \ddot{\chi}(t)=-m_2M_2^2 f(r), \ee
which does not allow a consistent nontrivial solution.
Thus in this model,
there is no stealth Schwarzschild solution 
with the ansatz $\phi_0= \chi(t)+\psi(r)$.

\subsubsection{Case $\phi_0(t,r)= \chi(t)\psi(r)$} 
\label{sec522}

Second, we consider the product separable ansatz 
$\phi_0=\chi(t)\psi(r)$,
for which Eq.~\eqref{t_eq} reads
\begin{align}
 \ddot{\chi}
=\frac{f^2\psi'^2}{\psi^2}
  \chi
-\frac{m_2M_2^2f}{\psi}.
\end{align} 
Since the left-hand side is independent on $r$, the right-hand side should be also independent of $r$.
In order for the right-hand side to be independent of $r$,
both $f\psi'/\psi$ and $f/\psi$
have to be constant in $r$.
However, they give $\psi \propto f$ and $\psi'={\rm const.}$
Clearly, these two requirements are not be compatible with $f(r)=1-2M/r$ for $M\neq 0$.
Thus, there is no stealth Schwarzschild solution
for the product separable case.

%%%%%%%%%
\subsection{Case $\phi_0=\phi_0(t,r,\theta,\varphi)$}
\label{sec53}

We then extend the analysis to more general cases $\phi_0=\phi_0(t,r,\theta,\varphi)$.
Since the equations are partial differential equations,
it is difficult to handle them explicitly. 
In the rest, we list the particular cases
where the stealth solutions cannot be obtained
for the model \eqref{g2conc}.
The same conclusions can also be obtained for the models  
\eqref{g3conc} and \eqref{g2g3conc}.

\subsubsection{Sum separable ansatz}
\label{sec531}

First,
we consider more general sum separable ansatz.
Following the discussion in Sec.~\ref{sec521},
the analysis for the ansatz $\phi_0=\chi(t)+\psi(r,\theta,\varphi)$
gives rise to no stealth Schwarzschild solution.

For the ansatz $\phi_0=\Phi(\varphi)+\psi(t,r,\theta)$,
Eq.~\eqref{vp_eq} yields
\begin{align} 
\label{g2_sum_phi} 
\Phi''(\varphi)
&=m_2 M_2^2 r^2\sin^2\theta.
\end{align}
which cannot be consistently satisfied,
when it is viewed as the equation for $\Phi$.
Thus, from the beginning we have to set $\Phi=0$,
and then the remaining Eqs.~\eqref{t_eq}, \eqref{r_eq}, and \eqref{th_eq} 
should be satisfied for a single function $\psi$,
and the consistent solution is only $\psi=0$,
namely the no-hair solution.

Similarly, 
for the ansatz $\phi_0=\Theta(\theta)+\psi(t,r,\varphi)$ 
Eq.~\eqref{th_eq} yields
\begin{align} 
\label{g2_sum_theta} 
\Theta''(\theta) 
=m_2 M_2^2 r^2+\frac{1}{\tan\theta\sin^2\theta\,\Theta'(\theta)}\psi_\varphi^2,
\end{align}
which also cannot be satisfied,
when it is viewed as the equation for $\Theta$.
Thus, from the beginning we have to set $\Theta=0$,
and then the remaining Eqs.~\eqref{t_eq}, \eqref{r_eq}, and \eqref{vp_eq} 
should be satisfied for a single function $\psi$,
and the consistent solution is only the no-hair solution $\psi=0$.

Finally, 
for the ansatz $\phi_0=R(r)+\psi(t,\theta,\varphi)$,
Eqs.~\eqref{t_eq}, \eqref{th_eq}, and \eqref{vp_eq}
do not depend on $R$ and its derivatives. 
Hence, the equations become those for $\psi(t,\theta,\varphi)$
with $r$-dependent coefficients,
for which the method of separation of variables does not work and hence they
cannot be consistently satisfied 
unless $\psi=0$,
to which the analysis in Sec.~\ref{sec441} applies.

Therefore, there is no stealth Schwarzschild solution for the sum separable ansatz about single coordinate.
Even for the sum separable cases about two coordinates, 
such as 
$\phi_0=\psi_1(t,r)+\psi_2(\theta,\varphi)$
and 
$\phi_0=\psi_1(t,\varphi)+\psi_2(r,\theta)$,
in general the four independent conditions \eqref{t_eq}--\eqref{vp_eq}
cannot be consistently satisfied
unless $\psi_1=\psi_2=0$.
Hence, we end up with the
no-hair solution.

\subsubsection{Product separable ansatz}
\label{sec532}

Let us consider the product separable ansatz for the stealth scalar field.
First, we consider the product separable ansatz for a single coordinate. 
For instance,
if we consider the ansatz of the scalar field $\phi_0=\chi(t)\psi (r,\theta,\varphi)$.
Eq.~\eqref{t_eq} becomes
\begin{align}
\label{eqchi}
\psi^2\ddot{\chi}
-f^2\psi'^2 \chi
-\frac{f\psi_\theta^2\chi}{r^2}
-\frac{f\psi_\theta^2\chi}{r^2\sin^2\theta}
+\frac{m_2M_2^2\psi}{f}
=0.
\end{align}
If Eq.~\eqref{eqchi} is viewed as an equation for $\chi(t)$,
it cannot be satisfied unless $\psi=0$,
since otherwise the coefficients depend on the other coordinates.
Similarly, for the ansatze 
$\phi_0=R(r)\psi (t,\theta,\varphi)$,
$\phi_0=\Theta(\theta)\psi (t,r,\varphi)$,
and 
$\phi_0=\Phi(\varphi)\psi (t,r,\theta)$,
the equations $R$, $\Theta$, and $\Phi$
cannot be satisfied unless $\psi=0$, respectively.
Thus, these product separable ansatz give the no-hair Schwarzschild solution.

Therefore, for these product separable ansatze about single coordinate,
there is no stealth Schwarzschild solution.
Even for the product separable cases about two coordinates, 
such as 
$\phi_0=\psi_1(t,r)\psi_2(\theta,\varphi)$
and 
$\phi_0=\psi_1(t,\varphi)\psi_2(r,\theta)$,
the four independent conditions \eqref{t_eq}--\eqref{vp_eq}
cannot be consistently satisfied, 
unless $\psi_1=\psi_2=0$.
Hence we end up with the no-hair solution.

%%%%%%%%%%%%%%%%%%%%%%%%%%%%%%%%%%%%%%%%%%%%%%%%%%%
\section{Conclusions}
\label{sec6}

In the present paper, we have found stealth Schwarzschild solutions
in the class of the shift symmetry breaking Horndeski theory~\eqref{action} with $G_4=\Mpl^2/2$,
where the propagation speed of gravitational waves coincide with the speed of light. 
Interestingly enough, these solutions exist only for shift-symmetry breaking theories, 
and one cannot obtain them for shift symmetric theories.

In Sec.~\ref{sec3}, we have derived the sufficient conditions 
for $G_2(\phi,X)$ and $G_3(\phi,X)$ to allow the Ricci-flat metric solutions  
and the stealth scalar field profile that does not affect the metric sector.
The covariant analysis in Sec.~\ref{sec3} applies 
any general Ricci-flat metric including the Schwarzschild and Kerr solutions, 
and a general scalar field profile.
The crucial point is that the analysis requires that 
the shift symmetry is broken in the scalar field sector, 
otherwise one would not obtain a nontrivial solution.
We provided \eqref{g2gen}, 
\eqref{g3gen}, and \eqref{g2g3gen} 
as general example theories satisfying the sufficient conditions.

In Secs.~\ref{sec4} and \ref{sec5}, 
we have applied the analysis of Sec.~\ref{sec3} to the Schwarzschild solution,
and considered the nontrivial scalar field profiles;
$\phi=\phi_0(r)$
which shares the symmetry with the metric functions,
and $\phi=\phi_0(t,r,\theta,\varphi)$
which does not share the symmetry with the metric functions,
respectively.
For the former case with $\phi=\phi_0(r)$ in Sec.~\ref{sec4}, 
we derived two types of stealth Schwarzschild solutions \eqref{sfc} and \eqref{sol0}.
The solution \eqref{sfc} is regular at the event horizon and 
exists for theories \eqref{g2conc}, \eqref{g3conc}, and \eqref{g2g3conc},
whereas the solution \eqref{sol0} is not regular at the event horizon and 
exists for the case $G_2(\phi,X)=0$ and $G_3(\phi,X)\neq 0$,
for which the conditions do not identify the specific form of $G_3(\phi,X)$.
Moreover, we investigated the linear perturbations about the solution~\eqref{sfc} and 
found that the kinetic term of the scalar mode identically vanishes. 
We also argued that this nature is universal for any stealth Schwarzschild solution
with time independent scalar field. 
On the other hand, we clarified in Sec.~\ref{sec5} that there is no stealth Schwarzschild solution for time dependent scalar field.

While the absence of the kinetic term of the scalar mode in the quadratic action indicates the strong coupling in the stealth Schwarzschild solution~\eqref{sfc} and requires nonlinear analysis, 
it is worthwhile to remark that 
the statement about the scalar mode is
{\it not} about the theories~\eqref{g2conc}, \eqref{g3conc}, and \eqref{g2g3conc} themselves,
but about the particular solutions \eqref{sfc}.
Indeed, since the theories satisfy the sufficient condition for the GR solution~\cite{Motohashi:2018wdq}, 
they also allow the Schwarzschild solution with constant scalar field profile, 
for which the analysis in Sec.~\ref{sec44} does not apply and independent analysis with $\phi=$~const.\ is required 
(See Sec.~IV~D and footnote 2 in \cite{Kobayashi:2014wsa}).

It is very interesting to consider other stealth Ricci-flat solutions, 
especially stealth Kerr solution,
which is more relevant for astrophysical applications.
We speculate that 
the absence of the kinetic term of the scalar mode in the quadratic action
crucially depends on the character of the scalar field.
If $\partial_\mu \phi$ is spacelike,
there would be a choice of time coordinate 
in which 
the kinetic term of the linear perturbations in the second order action vanishes
on the constant time hypersurface,
and the Cauchy problem is ill-posed.
As argued in Ref.~\cite{Motloch:2015gta},
even though the kinetic term in the second order action does not vanish 
for an alternative choice of the time coordinate,
the spacelike Cauchy surface that intersects with 
all the characteristic curves does not exist, 
as the ill-posedness of the Cauchy problem is diffeomorphism invariant.
On the other hand,
if $\partial_\mu \phi$ is timelike,
the scalar mode about the solution would have the kinetic term at the quadratic order.
The existence of an explicit stealth Ricci-flat solution
would depend on the choice of the metric solution and ansatz for the scalar field,
which is left for future work.

\acknowledgments{
We thank Shinji Mukohyama and Teruaki Suyama for fruitful discussions.
M.M.\ was supported by FCT-Portugal through Grant No.\ SFRH/BPD/88299/2012. 
H.M.\ was supported by JSPS KEKENHI Grant Nos.\ JP17H06359 and JP18K13565,
and the Centre for Cosmological Studies.
H.M.\ acknowledges 
Centro de Astrof\'{\i}sica e Gravita\c c\~ao and 
Institut Astrophysique de Paris 
for their hospitalities, where part of this work was completed.
}

\bibliography{ref-nohair}

\end{document}